\documentclass[prl,preprint,showpacs,showkeys,nofootinbib,tightenlines]{revtex4}

\usepackage{graphicx}  %
\usepackage{amsmath}

\def\vec#1{{\bf #1}}

\newcommand{\beq}{\begin{equation}}
\newcommand{\eeq}{\end{equation}}
\newcommand{\bqa}{\begin{eqnarray}}
\newcommand{\eqa}{\end{eqnarray}}

\def\mqo2{{\!\!\!}}

\begin{document}

\title{Universality Constraints on Three-Body Recombination \\
  for Cold Atoms: from $^4$He to $^{133}$Cs
}

\author{Eric Braaten and Daekyoung Kang}
\affiliation{Department of Physics,
         The Ohio State University, Columbus, OH\ 43210, USA}


\author{Lucas Platter}
\affiliation{Department of Physics and Astronomy, Ohio University,
        Athens, OH\ 45701, USA}

\date{\today}

\begin{abstract}
For atoms with large scattering length, the dependence of 
the 3-body recombination rate on the collision energy 
is determined by the scattering length and 
the Efimov 3-body parameters and can be expressed in terms 
of universal functions of a single scaling variable.
We use published results on the 3-body recombination rate 
for $^4$He atoms to constrain the universal functions. 
We then use those universal functions to calculate the
3-body recombination rate for other atoms with large scattering length
at nonzero temperature.
The constraints from the $^4$He results are strong if the scattering length 
is near the minimum of the 3-body recombination rate at threshold.
We apply our results to $^{133}$Cs atoms with a large 
positive scattering length, and compare them with 
experimental results from the Innsbruck group.
\end{abstract}

\smallskip
\maketitle

\section{Introduction}
Vitaly Efimov discovered in 1970 that 3-body systems of 
nonrelativistic particles with short-range interactions 
have remarkable universal properties if the S-wave 
scattering length $a$ is large compared to the range.
If $a = \pm \infty$, there are infinitely many 3-body 
bound states ({\it Efimov states} or {\it Efimov trimers})
with an accumulation point at the 
scattering threshold and a geometric spectrum \cite{Efimov70}: 
\begin{eqnarray}
E^{(n)}_T = (e^{-2\pi/s_0})^{n-n_*} \hbar^2 \kappa^2_* /m,
\label{kappa-star}
\end{eqnarray}
where $\kappa_*$ is the binding wavenumber of the branch 
of Efimov states labeled by $n_*$. This geometric spectrum 
is a signature of a {\it  discrete scaling symmetry}
with discrete scaling factor $e^{\pi/s_0}$ \cite{Braaten:2004rn}.
In the case of identical bosons, $s_0 \approx 1.00624$
and the discrete scaling factor is $e^{\pi/s_0} \approx 22.7$.
Efimov showed that the discrete scaling symmetry 
is also relevant for finite $a$ \cite{Efimov71}.
There is an infinite sequence of negative values of $a$ 
differing by $e^{\pi/s_0}$ and approaching $- \infty$ for which 
there is an Efimov trimer at the 3-atom scattering threshold.
There is also an infinite sequence of positive values of $a$ 
differing by $e^{\pi/s_0}$ and approaching $+ \infty$ for which 
there is an Efimov trimer at the atom-dimer scattering threshold. 
We will refer to the universal phenomena 
characterized by a discrete scaling symmetry that can occur 
in 3-body systems with a large scattering length as 
{\it Efimov physics}. Another example of Efimov physics 
that was discovered more recently is an infinite sequence
of positive values of $a$ differing by $e^{\pi/s_0}$ 
and approaching $+ \infty$ for which the 3-body recombination rate 
into the shallow dimer vanishes at the threshold \cite{NM-99,EGB-99,BBH-00}:
$a=(e^{\pi/s_0})^n a_{*0}$, where $a_{*0} \approx  0.32 \, \kappa_*^{-1}$.

In some systems, such as $^4$He atoms,
the universal aspects of Efimov physics are 
determined by two parameters: the scattering length $a$
and the Efimov parameter $\kappa_*$.
However, the situation is more complicated for the alkali atoms that are
used in most cold atom experiments as they form diatomic molecules
with many deep-lying bound states ({\it deep dimers}).
Efimov physics is modified by the existence of the deep dimers,
because Efimov trimers can decay into an atom and a deep dimer.
The deep dimers also provide additional 3-body recombination channels.
If there are deep dimers, the universal aspects of 
Efimov physics are determined by three parameters:
$a$, $\kappa_*$, and a parameter $\eta_*$ that determines 
the widths of Efimov trimers \cite{Braaten:2003yc}.

The first experimental evidence for Efimov physics 
has recently been presented by the Innsbruck group \cite{Grimm06}.
They carried out experiments with ultracold 
$^{133}$Cs atoms in the lowest hyperfine state,
using a magnetic field to control their scattering length. 
They observed a resonant enhancement in the 3-body recombination rate 
at $a \approx -850 \ a_0$ that can be attributed to an Efimov trimer 
near the 3-atom threshold.  At the temperature 10 nK, 
the loss rate as a function of $a$ can be fit 
rather well by the universal formula for zero temperature 
derived in Ref.~\cite{Braaten:2003yc}
with a width parameter $\eta_* = 0.06(1)$.
The Innsbruck group also observed a local minimum in the 
3-body recombination rate near $a \approx 210 \ a_0$ that might be
attributable to Efimov physics.  One complication is that 
$a$ is not large compared to the van der Waals length scale
for Cs atoms: $(mC_6/\hbar^2)^{1/4} \approx 200 \ a_0$.  
Universal predictions might not be quantitatively accurate 
for such a small value of $a$. Another 
complication is that the measurements were carried out at 200 nK.
At this temperature, there could be large corrections to the universal
predictions for zero temperature.

The 3-body recombination rate can be calculated at nonzero temperature 
by carrying out a thermal average of the 3-body recombination rate
as a function of the collision energy.  The universal predictions for 
nonzero collision energy have not yet been calculated.
There have been several previous efforts to calculate the 3-body 
recombination rate for atoms with large scattering length 
at nonzero temperature.
D'Incao, Suno, and Esry calculated the 3-body recombination rate 
at nonzero temperature for a simple model potential whose range 
and depth were tuned to give a single S-wave bound state and a 
large scattering length \cite{DSE-04}. They considered the qualitative effect of 
temperature on both the resonant enhancement in the case $a<0$ 
and on the local minimum in the case $a<0$.  
Jonsell \cite{Jonsell06} and Yamashita, Frederico, and Tomio
\cite{YFT06} considered the effects of temperature 
on the resonant enhancement in the case $a<0$. 
Jonsell used the adiabatic hyperspherical approximation to
calculate the 3-body elastic scattering rate as a function 
of the collision energy \cite{Jonsell06}.
Yamashita {\it et al.}~calculated the position and width of the Efimov 
resonance using a numerically exact method \cite{YFT06}. 
The two groups used their results to
estimate the 3-body recombination rate at nonzero temperature.
They were able to describe qualitatively the 
temperature dependence of the data from the Innsbruck experiment.
Neither of these calculations is a definitive universal prediction
for the 3-body recombination rate at nonzero temperature.

In this paper, we use published results on the 3-body 
recombination rate for $^4$He atoms to constrain the 
universal functions that determine its dependence 
on the collision energy.
Those universal functions are then used to calculate the 3-body 
recombination rate at nonzero temperature
for other atoms with large positive scattering length. 
The constraints from the $^4$He results are strong if the scattering length 
is near the minimum of the 3-body recombination rate at threshold.
We apply our results to $^{133}$Cs atoms with a large 
positive scattering length, and compare them with the
experimental results from the Innsbruck group.

\section{Three-body Recombination}

Three-body recombination is a 3-atom collision process in which 
two of the atoms bind to form a diatomic molecule (a dimer).
The 3-body recombination rate $R$ is a function of the momenta 
$\vec p_1$, $\vec p_2$, and $\vec p_3$ of the three incoming atoms.  
Galilean invariance implies that $R$ does not depend on the total 
momentum $\vec P_{\rm tot} = \vec p_1\, + \vec p_2\, + \vec p_3$.
It can therefore be expressed as a function 
of a pair of Jacobi momenta $\vec p_{12}$ and $\vec p_{3,12}$.
It is convenient to parameterize the Jacobi momenta in terms of 
the collision energy $E$, four angles describing the orientations 
of the Jacobi momenta, and a hyperangle $\alpha_3$ determined by the ratio 
of the magnitudes of the Jacobi momenta. 
We denote the 3-body recombination rate averaged over the angles 
and the hyperangle by $K_3(E)$:  
\begin{eqnarray}
K_3(E) 
= \langle R(\vec p_{12}\,, \vec p_{3,12}\,) 
\rangle_{\hat p_{12}, \, \hat p_{3,12}, \, \alpha_3} \,.
\label{K3-def}
\end{eqnarray}

If the scattering length $a$ is positive and large compared to the 
range of the interaction, one of the dimers 
that can be produced by the recombination process is the shallow dimer 
with binding energy 
\begin{eqnarray}
E_D = \hbar^2/(m a^2) \,.
\label{E-dimer}  
\end{eqnarray}
If there are deep dimers, they can be produced by the recombination
process for either sign of $a$.
The recombination rate can be decomposed 
into the contribution from the shallow dimer 
and the sum of the contributions from all the deep dimers:
\begin{equation}
K_3(E) =  K_{\rm shallow}(E) + K_{\rm deep}(E)  \,.
\label{K3}
\end{equation}
The 3-body recombination rate into the shallow dimer 
can be further decomposed into contributions from the channels 
in which the total orbital angular momentum of the three atoms
has definite quantum number $J$:
\begin{equation}
K_{\rm shallow}(E) = \sum_{J=0}^\infty K^{(J)}(E)  \,.
\label{K-shallow}
\end{equation}
The scaling behavior near the scattering threshold for each of the
angular momentum contributions is known \cite{SEGB02}:
$K^{(J)}(E) \sim E^{\lambda_J}$,
where $\lambda_J = 0$, 3, 2, and 3 for $J=0$, 1, 2, and 3.
At the scattering threshold $E=0$, only the $J=0$ term is nonzero.

In a gas of atoms with number density $n_A$,
the rate of decrease in the number density due to 3-body recombination
defines an experimentally measurable loss rate constant $L_3$: 
\begin{eqnarray}
\frac{d \ }{d t} n_A 
= - L_3 \, n_A^3 \,.
\label{dn-lost}
\end{eqnarray}
The event rate constant $\alpha$ for 3-body recombination
is defined so that the number of recombination 
events per time and per volume is $\alpha \, n_A^3$.  
If the number of atoms lost from the system per recombination event 
is $n_{\rm lost}$, the rate constant is $L_3 = n_{\rm lost} \, \alpha$.
The binding energy of the dimer is released through the  
kinetic energies of the recoiling atom and dimer. 
If their kinetic energies are large enough that the
atom and dimer both escape 
from the system and if they don't undergo further collisions 
before escaping, 
then $n_{\rm lost} = 3$.
On the other hand, if they both remain in the system 
and if we regard the dimer as a distinct chemical species, 
then $n_{\rm lost} = 2$.

In an ensemble of identical bosons with number density $n_A$
and 3-particle momentum distribution 
$f(\vec p_1\,,\vec p_2\,,\vec p_3\,)$, the event rate 
per volume and per time is
\begin{eqnarray}
\alpha \, n_A^3 = 
\frac{1}{6} \int R(\vec p_{12}\,, \vec p_{3,12}\,) \,
f(\vec p_1\,,\vec p_2\,,\vec p_3\,) \,
\frac{d^3 p_1}{(2 \pi)^3} \frac{d^3 p_2}{(2 \pi)^3} \frac{d^3 p_3}{(2 \pi)^3}\,.
\label{Rfint}  
\end{eqnarray}
The factor of $1/6$ compensates for overcounting the states of the 
three identical particles by integrating over the entire range 
of the three momenta.
The 3-particle momentum distribution is normalized so that the number of
3-particle states per volume-cubed is $n_A^3/3!$:
\begin{eqnarray}
\frac{1}{6} \int 
f(\vec p_1\,,\vec p_2\,,\vec p_3\,) \,
\frac{d^3 p_1}{(2 \pi)^3} \frac{d^3 p_2}{(2 \pi)^3} \frac{d^3 p_3}{(2 \pi)^3}
= \frac{n_A^3}{6} \,.
\label{f-norm}  
\end{eqnarray}
Dividing Eq.~(\ref{Rfint}) by Eq.~(\ref{f-norm}), we get an expression 
for $\alpha$ as a ratio of two integrals.

If the system is in thermal equilibrium at temperature $T$, 
the 3-particle momentum distribution 
$f(\vec p_1\,,\vec p_2\,,\vec p_3\,)$ is the product of three
Bose-Einstein distributions.  If $T$ is large compared to the 
critical temperature for Bose-Einstein condensation,
$k_B T_c \gg 3.3 \, \hbar^2 n_A^{2/3}/m$,
the 3-particle momentum distribution can be approximated by 
a Boltzmann distribution proportional to $\exp(-E_{\rm tot}/(k_B T))$,
where $E_{\rm tot}$ is the total energy of the three atoms.  
The total energy is the sum of the center-of-mass energy 
and the collision energy $E$: 
$E_{\rm tot} = P_{\rm tot}^2/(6m) + E$.
The Boltzmann factor is therefore the product of a Boltzmann
factor for the center-of-mass energy and a Boltzmann
factor for the collision energy.
Since the recombination rate does not depend on 
the total momentum $P_{\rm tot}$,
the event rate constant $\alpha(T)$  reduces to 
the Boltzmann average over the collision energy $E$:
\begin{eqnarray}
\alpha(T) = 
\frac{\int_0^\infty dE \, E^2 \, e^{-E/(k_B T)} \, K_3(E)}
    {6 \int_0^\infty dE \, E^2 \, e^{-E/(k_B T)}} \,.
\label{alpha-T}  
\end{eqnarray}
The weight factor $E^2$ comes from using hyperspherical variables 
for the Jacobi momenta.  The integral in the denominator 
can be evaluated analytically to give $2 (k_B T)^3$.

\section{Universal behavior}

In this section and in the subsequent two sections,
we consider atoms for which the only diatomic molecule
is the shallow dimer and there are no deep dimers.
The total 3-body recombination rate $K_3(E)$ 
is therefore equal to $K_{\rm shallow}(E)$.
An example of an atom with a large positive scattering length 
and no deep dimers is $^4$He.  In this section,
we summarize the universal information that is known 
on the recombination rate $K_3(E)$ for this case. 

At the scattering threshold $E=0$, only the $J=0$ term 
in the decomposition of $K_{\rm shallow}(E)$ into orbital 
angular momentum channels in Eq.~(\ref{K-shallow}) is nonzero.
An analytic expression for this term has recently been derived
\cite{Petrov-octs,MOG05}:
\begin{eqnarray}
K^{(0)}(E=0) = 
\frac{768 \pi^2 (4 \pi - 3 \sqrt{3}) \sin^2 [s_0 \ln (a/a_{*0})]}
    {\sinh^2(\pi s_0) + \cos^2 [s_0 \ln (a/a_{*0})]} \,
\frac{\hbar a^4}{m} \,.
\label{K0-analytic}
\end{eqnarray}
The parameter $a_{*0}$ differs from $\kappa_*^{-1}$ by a multiplicative 
constant that is known only to a couple digits of accuracy:
$a_{*0}  \approx 0.32 \, \kappa_*^{-1}$.
The coefficient of $\hbar a^4/m$ in Eq.~(\ref{K0-analytic})
is a log-periodic function 
of $a/a_{*0}$ that oscillates between zero and 
a maximum value $C_{\rm max}$ as a function of $\ln(a)$.
The analytic expression for the maximum value is%
\footnote{This constant differs from the constant $C_{\rm max}$
in Ref.~\cite{BH02} by a factor of 6.} 
\begin{eqnarray}
C_{\rm max} = 
\frac{768 \pi^2 (4 \pi - 3 \sqrt{3})}{\sinh^2(\pi s_0)} \,.
\label{K-max}
\end{eqnarray}
Its numerical value is $C_{\rm max} \approx 402.7$.
The expression in Eq.~(\ref{K0-analytic}) has zeroes when 
$a$ is $( e^{\pi/s_0})^n \, a_{*0}$, where $n$ is an integer.
Since $\sinh^2(\pi s_0) \approx 139$ is so large,
$K^{(0)}(0)$ can be approximated 
with an error of less than 1\% of $C_{\rm max}\hbar a^4/m$ by
\begin{equation}
K^{(0)}(E=0) \approx 
C_{\rm max} \sin^2 [s_0 \ln (a/a_{*0}) ] \, \hbar a^4/m \,.
\label{alpha-sh}
\end{equation}
This approximate functional form of the rate constant was first deduced
in Refs.~\cite{NM-99,EGB-99}.
The coefficient $C_{\rm max}$ and the relation between $a_{*0}$ 
and $\kappa_*$ was calculated numerically in Refs.~\cite{BBH-00,BH02}. 
The maxima of the coefficient of $\hbar a^4/m$ in
$K^{(0)}(0)$ in Eq.~(\ref{alpha-sh}) occur when 
$a$ is $( e^{\pi/s_0} )^n \, 4.76 \, a_{*0}$.
The maxima of 
$K^{(0)}(0)$ in Eq.~(\ref{alpha-sh}) occur when 
$a$ is $( e^{\pi/s_0} )^n \, 14.3 \, a_{*0}$.

If the collision energy $E$ is small compared to the natural energy scale
$\hbar^2/(m \ell^2)$ set by the range $\ell$, the 3-body recombination rate
$K_{\rm shallow}(E)$ is a universal function of the  
collision energy $E$, the scattering length $a$, and the 
Efimov parameter $a_{*0}$.
The calculation of the rate can be reduced to the 
calculation of universal functions of a scaling variable $x$ 
defined by
\begin{equation}
x = (m a^2 E/\hbar^2)^{1/2} \,.
\label{x-def}
\end{equation}
Only the $J=0$ term in $K_{\rm shallow}(E)$ depends on $a_{*0}$,
and that dependence is log-periodic with discrete scaling factor 
$e^{\pi/s_0} \approx 22.7$.  The constraints of universality are 
therefore particularly simple for $J \ge 1$. 
These terms are determined by a single universal function for each $J$:
\begin{equation}
K^{(J)}(E) = f_J(x) \, \hbar a^4/m \,.
\label{alpha-J:uni}
\end{equation}

The constraints of universality are more intricate for $J=0$.  
They can be deduced from {\it Efimov's radial law}, which follows 
from unitarity and from the fact that there is only
one attractive adiabatic hyperspherical potential in the 
scale-invariant region of the hyperradius $R$: $\ell \ll R \ll a$.
Efimov's radial law implies that the $J=0$ term in the 3-body 
recombination rate must have the form  \cite{Braaten:2004rn}
\begin{equation}
K^{(0)}(E) = \frac{k}{x^4} 
\left| s_{32}(x) 
+ \frac{s_{31}(x) s_{12}(x) e^{2 i s_0 \ln(a/a_{*0})}}
      {1 - s_{11}(x) e^{2 i s_0 \ln(a/a_{*0})}}
\right|^2 \frac{\hbar a^4}{m} \,,
\label{K0:rl}
\end{equation}
where $k$ is a constant and the functions
$s_{ij}(x)$ are entries of a symmetric $3 \times 3$ unitary matrix.
The values of $s_{11}(x)$ and $s_{12}(x)$ at the threshold
are known analytically up to phases \cite{MOG05}:
\begin{subequations}
\begin{eqnarray}
s_{11}(0) &=& e^{- 2 \pi s_0} \, e^{2 i \delta_1}\,,
\label{s11}
\\
s_{12}(0) &=& \sqrt{1 - e^{- 4 \pi s_0}} \, e^{2 i \delta_2} \,.
\label{s12}
\end{eqnarray}
\label{s1112}
\end{subequations}
The entries $s_{32}(x)$ and $s_{31}(x)$ go to zero as $x^2$ as
the threshold is approached.  Their ratio at the threshold 
is also known analytically \cite{MOG05}: 
\begin{equation}
s_{31}(x)/s_{32}(x)  \longrightarrow 
\sqrt{\coth(\pi s_0)} \, e^{2 i (\delta_1 -\delta_2)} 
\hspace{1cm} {\rm as \ } x \to 0\,.
\end{equation}
This limiting behavior together with the analytic values in 
Eqs.~(\ref{s1112}) are sufficient to deduce 
the analytic result in Eq.~(\ref{K0-analytic}) up to a multiplicative
normalization constant provided that $\delta_1 = \pm \pi/2$.

\section{Analysis of $^4$He atoms}

$^4$He atoms are a classic example of atoms with a large 
positive scattering length.  There are several modern potentials 
that are believed to describe the interactions of $^4$He atoms 
with low energy accurately.  The only one for which 
the 3-body recombination rate has been calculated is the
HFD-B3-FCI1 potential \cite{AJM95}.  The scattering length 
for the HFD-B3-FCI1 potential is $a = 172 \, a_0$.
This is much larger than either the effective range $r_s = 14 \, a_0$ 
or the van der Waals length scale $\ell_{\rm vdW} = 10.2 \, a_0$.  
The large scattering length $a$ implies that $^4$He atoms
have universal properties that are determined by $a$.
These properties can be calculated as expansions in powers 
of the range divided by $a$.
We refer to the leading order in this expansion as the 
{\it scaling limit} or the {\it zero-range limit}.
A thorough analysis of the universal properties of $^4$He atoms
in the scaling limit has been presented by Braaten and Hammer \cite{BH02}.
Various 3-body observables for $^4$He atoms have been calculated by 
Platter and Phillips to next-to-next-to-leading order in the expansion 
in powers of the range divided by $a$ \cite{Platter:2006ev}. 
The agreement with numerical results from exact solutions of the 3-body 
Schr\"odinger equation is impressive \cite{RY00,Ro03}.

The $^4$He potential supports a diatomic molecule 
with a single energy level.  The binding energy of this shallow dimer 
in the HFD-B3-FCI1 potential is $E_D = 1.600$ mK \cite{private}.  
Experience has shown that 
the universal predictions for 3-body observables are 
significantly more accurate if the dimer binding energy 
is used as the 2-body input instead of $a$ \cite{BH02}.
Using the universal expression for the dimer binding energy 
in Eq.~(\ref{E-dimer}), we obtain the scattering length%
\footnote{A convenient conversion constant for $^4$He atoms
$\hbar^2/m = 43.2788 \, {\rm K} \, a_0^2 $.}
\begin{equation}
a_{\rm He} = 164.5 \, a_0 \,.
\end{equation}

\begin{figure}[htb]
\centerline{\includegraphics*[width=12cm,angle=0,clip=true]{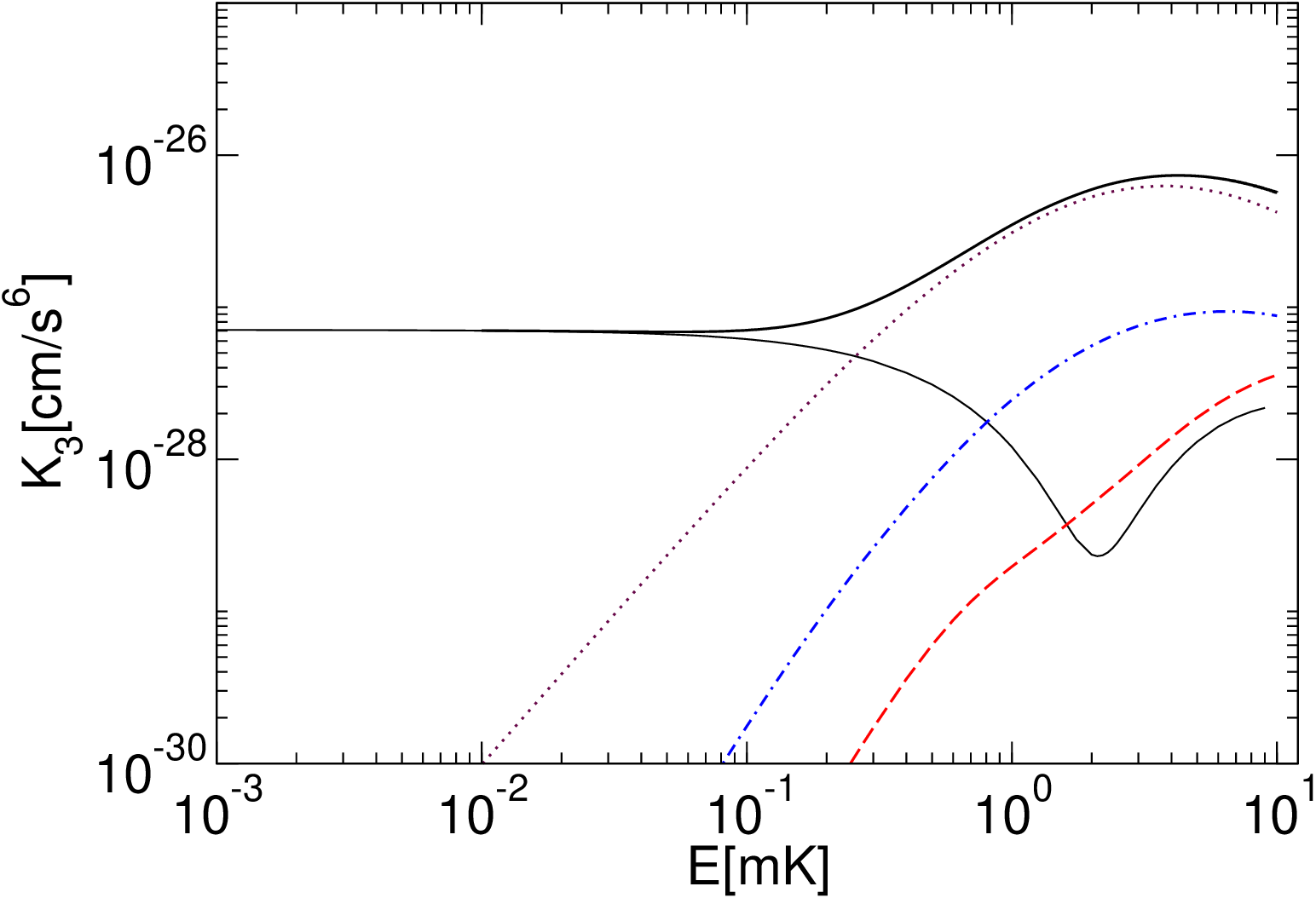}}
\vspace*{0.0cm}
\caption{The 3-body recombination rate $K_3(E)$ 
for $^4$He atoms (in units of ${\rm cm}^6/{\rm s}$) 
as a function of the collision energy $E$ (in units of mK)
from Ref.~\cite{SEGB02}.  The four lowest curves at $E=10$~mK 
starting from the bottom are the contributions
to $K_3(E)$ from $J=0$, 1, 3 and 2 (thin solid, dashed, dotted
and dot-dashed curve, respectively). The solid curve at the top 
is their sum.
}
\label{fig:alpha-He}
\end{figure}

The 3-body recombination rate $K_3(E)$ for $^4$He atoms 
interacting through the HFD-B3-FCI1 potential \cite{AJM95}
has been calculated as a function of the collision energy
from the threshold to 10 mK \cite{SEGB02}.
The results were separated into the contributions from
$J=0$, 1, 2, and 3.  The individual contributions and their sum 
are shown in Fig.~\ref{fig:alpha-He}.
We can determine the value of $a_{*0}$ for $^4$He atoms 
interacting through the HFD-B3-FCI1 potential from the value 
of the 3-body recombination loss rate at threshold:  
$K_3(0) =  7.10 \times 10^{-28}$ cm$^6$/s.  Inserting 
$a_{\rm He} = 164.5 \, a_0$ in Eq.~(\ref{K0-analytic})
and solving for $a_{*0}$, we obtain 
\begin{equation}
a^{\rm He}_{*0} = 143.1 \ a_0.
\label{astar0He}
\end{equation}
The result is approximately equal to $1.150 \,a_{\rm He}$.
The near equality between $a^{\rm He}_{*0}$ and $a_{\rm He}$
reflects the fact that $K_3(0)$ for $^4$He is much smaller 
than the maximum value 
$C_{\rm max} \hbar a_{\rm He}^4/m =  3.67 \times 10^{-26}$ cm$^6$/s
allowed by universality, which implies that 
$a/a_{*0}$ is close to a zero of the sine function in 
the numerator of Eq.~(\ref{alpha-sh}).
The value of $a_{*0}$ for $^4$He can also be determined 
from the binding energy of the excited $^4$He trimer, which is
$E_3^{(1)} = 2.62$ mK for the HFD-B3-FCI1 potential \cite{private}.
The resulting value, $a_{*0} \approx 146 ~ a_0$, is consistent 
with the value in Eq.~(\ref{astar0He}).  Its accuracy is limited 
by the accuracy to which 
the numerical value of the product $a_{*0} \kappa_* \approx 0.32$
is known.

We can use the $^4$He results calculated in Ref.~\cite{SEGB02}
to determine the universal functions in the contributions to the
3-body recombination rate for $J=1$, 2, and 3.
By fitting the results of Ref.~\cite{SEGB02} for $J=1$, 2, and 3,
the universal functions $f_J(x)$ defined by Eq.~(\ref{alpha-J:uni})
can be determined in the range $0 < x < 2.5$. 
The behavior of these universal functions near the threshold are
\begin{subequations}
\begin{eqnarray}
f_1(x) &\approx& 3.43 \, x^6,
\\
f_2(x) &\approx& 261 \, x^4,
\label{f2-x}
\\
f_3(x) &\approx& 90.3 \, x^6.
\end{eqnarray}
\end{subequations}

The results of Ref.~\cite{SEGB02} for $J=0$ do not contain enough
information to determine all the independent universal functions 
in the expression in Eq.~(\ref{K0:rl}) that comes from 
Efimov's radial law.  However, if we make a few plausible 
simplifying assumptions, the universal functions for $J=0$ 
can be strongly constrained.
According to Eq.~(\ref{s11}), the absolute value of the function 
$s_{11}(x)$ is approximately 0.002 at the threshold $x=0$.
Our first simplifying assumption is that this function remains 
small compared to 1 and can be neglected for $x < 2.5$.
The expression for the 3-body recombination rate in Eq.~(\ref{K0:rl})
can then be reduced to 
\begin{eqnarray}
K^{(0)}(E) &\approx& C_{\rm max} 
\Big| \big( \sin[s_0 \ln(a/a_{*0})] (1 + h_1(x)) 
        + \cos[s_0 \ln(a/a_{*0})] h_2(x) \big)
\nonumber
\\
&& \hspace{1cm}
+ i \big( \sin[s_0 \ln(a/a_{*0})] h_3(x) 
        + \cos[s_0 \ln(a/a_{*0})] h_4(x) \big) \Big|^2 \frac{\hbar a^4}{m} \,,
\label{K3-num}
\end{eqnarray}
where the functions $h_i(x)$ are real-valued and vanish at $x=0$.
In the case of $^4$He, the coefficients in Eq.~(\ref{K3-num}) are
\begin{subequations}
\begin{eqnarray}
\sin[s_0 \ln(a_{\rm He}/a^{\rm He}_{*0})] &=& 0.140, 
\\
\cos[s_0 \ln(a_{\rm He}/a^{\rm He}_{*0})] &=& 0.990.
\end{eqnarray}
\end{subequations}

A dramatic feature of the $J=0$ contribution to $K_3(E)$ for $^4$He 
in Fig.~\ref{fig:alpha-He} is the deep local minimum at the collision 
energy $E_{\rm min} = 2.1$ mK.  The rate at the minimum is
$K^{(0)}(E_{\rm min}) =  2.3 \times 10^{-29}$ cm$^6$/s. 
The rate decreases by more than an order of magnitude as $E$ goes from
0.6 mK to 2.1 mK and then it increases by almost an order of magnitude 
as $E$ goes from 2.1 mK to 10 mK.
This dramatic behavior can be reproduced by slowly varying universal
functions if the real part of the amplitude
inside the absolute value sign in Eq.~(\ref{K3-num}) changes sign 
at $E_{\rm min}$ and if the imaginary part of the amplitude
is small compared to the real part except near $E_{\rm min}$. 
Our second simplifying assumption is therefore that the imaginary part 
of that amplitude can be neglected for  $x < 2.5$. 
Thus the expression for the rate that follows from our two 
simplifying assumptions is 
\begin{eqnarray}
K^{(0)}(E) &=& C_{\rm max} 
\big| \sin[s_0 \ln(a/a_{*0})] (1 + h_1(x)) 
        + \cos[s_0 \ln(a/a_{*0})] h_2(x) \big|^2  \hbar a^4/m\,.
\label{K3-app}
\end{eqnarray}
We expect this to be a good approximation 
except near those values of $a$ and $E$ at which it vanishes,
which in the case of $^4$He is at $E = 2.1$ mK.

The approximation for $K^{(0)}(E)$ in Eq.~(\ref{K3-app})
still involves two unknown functions $h_1(x)$ and $h_2(x)$.
The $J=0$ results for $^4$He in Ref.~\cite{SEGB02} can be used 
to determine only a specific linear  combination of those
two functions:
\begin{equation}
h_{\rm He}(x) = 0.140 \, h_1(x) + 0.990 \, h_2(x) \,.
\label{hHE}
\end{equation}
To determine this linear combination, we fit the $J=0$ results  
of Ref.~\cite{SEGB02} using the expression  
\begin{eqnarray}
K^{(0)}_{\rm He}(E) &=& C_{\rm max} 
\big[ 0.140 + h_{\rm He}(x) \big]^2 \hbar a_{\rm He}^4/m
+ K^{(0)}_{\rm He}(E_{\rm min}) \left( 2E/(E_{\rm min} + E) \right)^2 \,.
\label{K3-He}
\end{eqnarray}
The last term in Eq.~(\ref{K3-He}) is a simple model for the term
in $K^{(0)}(E)$ that comes from the square of the imaginary part of 
the amplitude in Eq.~(\ref{K3-num}).  
It has the correct value at the threshold, where it must vanish,
and at $E_{\rm min} = 2.1$ mK.
This model affects the fit for $h_{\rm He}(x)$ 
only in the region very close to
$E_{\rm min}$.  The resulting function $h_{\rm He}(x)$
vanishes at $x=0$, decreases monotonically to $-0.135$ 
at $x = 1.15$, 
and then decreases further to $-0.206$ at $x = 2.5$.

\section{Other systems without deep dimers}

If the 3-body recombination rate had been calculated for another 
system of identical bosons with a large positive scattering length,
such as $^4$He atoms with a potential other than HFD-B3-FCI1,
we would be able to use those results to determine 
a different linear combination of $h_1(x)$ and $h_2(x)$
than in Eq.~(\ref{hHE}).
We would then know the two independent functions $h_1(x)$ and $h_2(x)$,
and we would be able to calculate $K^{(0)}(E)$ for any system 
of identical bosons 
with large positive scattering length.  Although we cannot determine 
$h_1(x)$ and $h_2(x)$ from the results for $^4$He, we can put 
constraints on them if assume that $h_{\rm He}(x)$ is not 
unnaturally small because of a near 
cancellation between the two terms in Eq.~(\ref{hHE}).
The constraint in Eq.~(\ref{hHE}) on 
$h_1(x)$ and $h_2(x)$ can be implemented by expressing them in
the form
\begin{subequations}
\begin{eqnarray}
h_1(x) &=& \frac{t(x)}{0.140} h_{\rm He}(x),
\\
h_2(x) &=& \frac{1-t(x)}{0.990} h_{\rm He}(x),
\end{eqnarray}
\label{h12}
\end{subequations}
where $t(x)$ is an unknown function of $x$. 
If $t(x)$ is in the range $0 < t(x) < 1$,
the two terms in Eq.~(\ref{hHE}) contribute to $h_{\rm He}(x)$
with the same sign.  If $t(x)$ is in the range 
\begin{equation}
- 2 < t(x) < + 3 \,,
\label{t-range}
\end{equation}
the two terms in Eq.~(\ref{K3-He}) can have opposite signs, 
but the absolute value of their sum can not be smaller 
than that of the larger of the two terms by more than a factor of 3.
This range is consistent with our assumption that $h_{\rm He}(x)$
is not unnaturally small.
Given this assumption, we can generate a range of 
predictions for the 3-body recombination rate 
for other systems with large scattering lengths by varying 
$t(x)$ in the range given by Eq.~(\ref{t-range}).

\begin{figure}[htb]
\centerline{\includegraphics*[width=12cm,angle=0,clip=true]{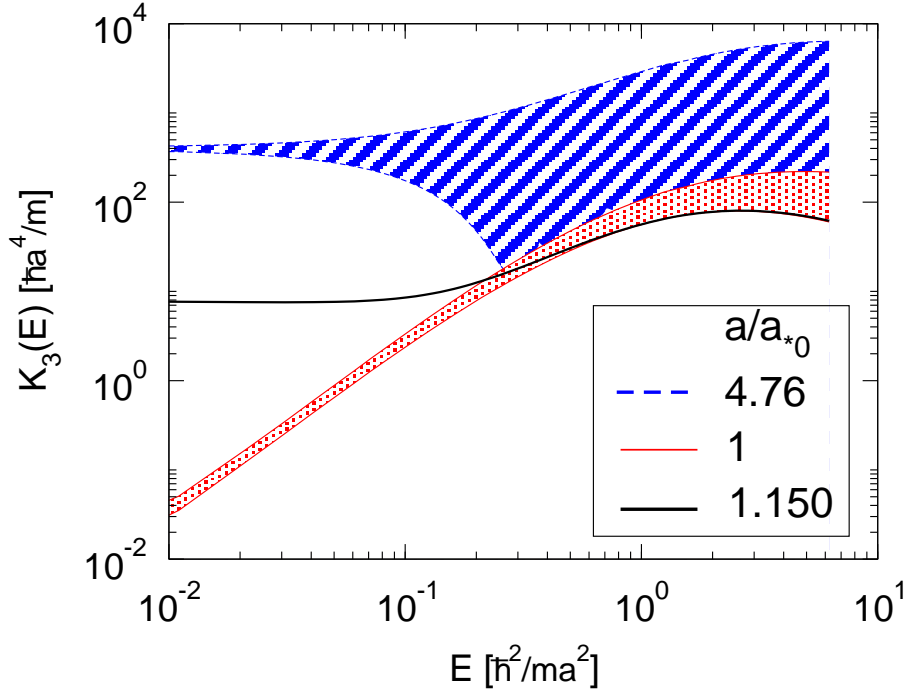}}
\vspace*{0.0cm}
\caption{The 3-body recombination rate $K_3(E)$ 
for atoms with no deep dimers and a large scattering length.
The rate $K_3(E)$ (in units of $\hbar a^4/m$) 
is shown as a function of the collision energy $E$ 
(in units of $\hbar^2/(m a^2)$)
for various values of the ratio $a/a_{*0}$:
1 (solid lines), 
1.150 (heavy solid line), and
4.76 (dashed lines). 
The band of predictions for each value of $a/a_{*0}$ corresponds to  
the range $-2 < t < +3$ for the variable $t$ in Eq.~(\ref{h12}).}
\label{fig:Kshallow-E}
\end{figure}

The 3-body recombination rate $K_3(E)$ is shown in 
Fig.~\ref{fig:Kshallow-E} as a function of the collision energy $E$ 
for three values of the ratio $a/a_{*0}$ of the scattering length 
to the Efimov parameter:
$a/a_{*0} = 1$, for which $K_{\rm shallow}(0)=0$,
$a/a_{*0} = 1.150$, which corresponds to $^4$He,
and $a/a_{*0} \approx 4.76$, 
for which $K_{\rm shallow}(0) = C_{\rm max} \hbar a^4/m$.
The rate $K_3(E)$ is the sum of the $J=0$ term in Eq.~(\ref{K3-app})
and the $J=1$, 2, and 3 terms in Eq.~(\ref{alpha-J:uni}).
The rate for each value of $a/a_{*0}$ is shown 
as an error band that corresponds to varying the parameter $t$ 
defined by Eqs.~(\ref{h12}) over the range in Eq.~(\ref{t-range}).
Since the error band comes only from the $J=0$ term, it always
remains above the contribution from the $J=2$ term.
In the $^4$He case, $a/a_{*0} = 1.150$, the band reduces to a 
curve given by the heavy solid line.
For $a/a_{*0} = 1$, the rate is dominated by the $J=2$ term
and the error band is always fairly narrow. 
It extends above the $J=2$ curve by a factor that is about 1.5 
for $E < 0.25 \, E_D$ and increases to about 3.6 at $E= 6.25 E_D$.
For $a/a_{*0} = 4.76$, the error band is narrow for $E < 10^{-2} \, E_D$,
but its width increases quickly with $E$.  
For $E > 0.25 \, E_D$, the error band extends above the $J=2$ curve 
by about two orders of magnitude.

\begin{figure}[htb]
\centerline{\includegraphics*[width=12cm,angle=0,clip=true]{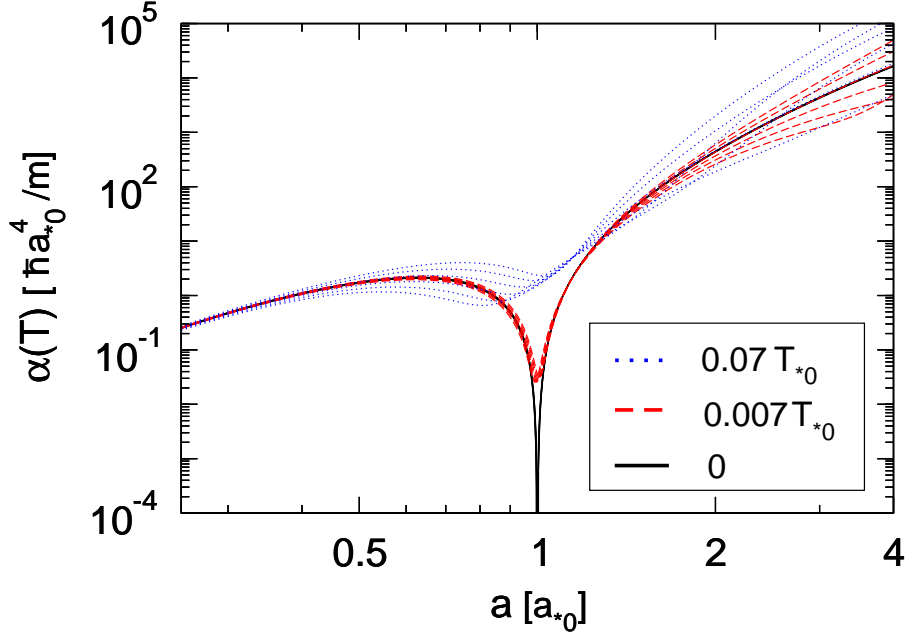}}
\vspace*{0.0cm}
\caption{The 3-body recombination event rate constant $\alpha(T)$ 
for atoms with no deep dimers and a large scattering length.
The rate constant $\alpha(T)$ (in units of $\hbar a_{*0}^4/m$) 
is shown as a function of $a$ 
for fixed $a_{*0}$ and several values of the temperature $T$:
$T = 0$ (solid line), 
$0.007 \, T_{*0}$ (dashed line), 
$0.07 \, T_{*0}$ (dotted line), 
where $k_B T_{*0} = \hbar^2/(m a_{*0}^2)$.
The 6 curves for each temperature $T$ correspond to 
integer values of $t$ from $-2$ to $+3$.}
\label{fig:alphashallow-T}
\end{figure}

Given a range of predictions for $K_3(E)$, we can obtain
a range of predictions for the 3-body recombination event rate 
constant $\alpha(T)$ given by Eq.~(\ref{alpha-T}).  
Since the results in Ref.~\cite{SEGB02} only allow us to determine 
the function $h_{\rm He}(x)$ in the region  $x < 2.5$,
the integral in the numerator of Eq.~(\ref{alpha-T}) can only be 
evaluated up to the energy $E_{\rm max} = 6.25 \, E_D$. 
As long as $T < 0.74 \, E_D$, more than 99\% of the weight 
of the normalizing integral in the denominator of Eq.~(\ref{alpha-T}) 
is in the region $x < 2.5$. 
In Fig.~\ref{fig:alphashallow-T}, $\alpha(T)$ is shown as a function of $a$
for fixed $a_{*0}$ and several values of $T$.
The range of predictions 
is shown by plotting 6 curves corresponding to integer values 
of the variable $t$ defined in Eq.~(\ref{h12}) ranging from $-2$ to $+3$.
The widths of the error bands reduce to zero at the $^4$He value 
$a=1.150 \, a_{*0}$.  The error bands remain narrow in a region 
that includes the 
value $a =  a_{*0}$ at which $K_{\rm shallow}(E=0)$ vanishes.
We therefore obtain almost unique predictions for $\alpha(T)$
near its local minimum as a function of $a$.
The depth of the minimum decreases as the temperature $T$ increases.

\section{Effects of deep dimers}

In this section and in the subsequent section, 
we consider atoms that have deep dimers.
In the scaling (or zero-range) limit, 
the cumulative effect of all the deep dimers on Efimov physics 
can be taken into account rigorously through one
additional parameter $\eta_*$ \cite{Braaten:2003yc}. 
The effects of Efimov physics can be dramatic only if 
$\eta_*$ is much less than 1. If the universal 
expression for a scattering amplitude for the case of no deep dimers 
is known as an analytic function of $a_{*0}$, 
the corresponding result for a system with deep
dimers can be obtained without any additional calculation simply
by making the substitution   
\begin{eqnarray}
\ln a_{*0} \longrightarrow \ln a_{*0} - i \eta_*/s_0 \,.
\label{logsub}
\end{eqnarray}
Making this substitution in the factor $\sin [s_0 \ln(a/a_{*0})]$ 
in Eq.~(\ref{alpha-sh}), the resulting expression for the 
recombination rate at threshold into the shallow dimer is
\begin{equation}
K_{\rm shallow}(E=0) \approx 
C_{\rm max} \left( \sin^2 [s_0 \ln (a/a_{*0}) ] + \sinh^2 \eta_* \right) \, 
\hbar a^4/m \,.
\label{K-shallow:eta}
\end{equation}
This expression shows that one effect of the deep dimers 
is to eliminate the zeroes of $K_{\rm shallow}(0)$.
Making this substitution
in Eq.~(\ref{K3-app}), the resulting expression 
for the $J=0$ contribution to $K_{\rm shallow}(E)$ is
\begin{eqnarray}
K^{(0)}(E) &=& C_{\rm max} \Big[ 
\cosh^2 \eta_* \big( \sin[s_0 \ln(a/a_{*0})] (1 + h_1(x)) 
        + \cos[s_0 \ln(a/a_{*0})] h_2(x) \big)^2
\nonumber
\\
&& 
\hspace{0.25cm}
+ \sinh^2 \eta_* \big( \cos[s_0 \ln(a/a_{*0})] (1 + h_1(x)) 
        - \sin[s_0 \ln(a/a_{*0})] h_2(x) \big)^2 \Big] 
\frac{\hbar a^4}{m} \,.
\label{K3-app:deep}
\end{eqnarray}

If there are deep dimers, there is an additional contribution 
$K_{\rm deep}(E)$ from 3-body recombination rate into deep dimers.
A generalization of Efimov's radial law implies that $K_{\rm deep}(E)$
must have the form \cite{Braaten:2004rn}
\begin{equation}
K_{\rm deep}(E) = 
\frac{k |s_{31}(x)|^2 (1 - e^{-4\eta_*})}
    {x^4\left| 1 - s_{11}(x) e^{2 i s_0 \ln(a/a_{*0})-2 \eta_*} \right|^2}
\frac{\hbar a^4}{m} \,,
\label{K-deep:rl}
\end{equation}
where $k$ is the same constant and $s_{11}(x)$ and 
$s_{31}(x)$ are the same entries of a symmetric $3 \times 3$ unitary matrix
that appear in Eq.~(\ref{K0:rl}).
We can obtain an analytic expression for the value at the threshold
$E=0$ by using the value of $s_{11}(0)$  given in Eq.~(\ref{s11})
and the limiting behavior of $s_{31}(x)$ as $x \to 0$: 
\begin{eqnarray}
K_{\rm deep}(E=0) = 
\frac{C_{\rm max} \sinh(2\pi s_0) \sinh(2\eta_*)}
    {4 \left( \sinh^2(\pi s_0 + \eta_*) + \cos^2[s_0 \ln(a/a_{*0})] \right) }\,
\frac{\hbar a^4}{m} \,.
\label{K-deep:eta}
\end{eqnarray}
Since $\sinh^2(\pi s_0+ \eta_*) > 139$ is so large,
$K_{\rm deep}(0)$ can be approximated 
with an error of less than 1\% by omitting the $\cos^2$ term 
in the denominator of Eq.~(\ref{K-deep:eta}).  To within the same 
accuracy, we can replace $\sinh(2\pi s_0)$ and $\sinh(\pi s_0 + \eta_*)$
by exponentials to get the simpler expression
\begin{eqnarray}
K_{\rm deep}(E=0) \approx 
\frac{(1 - e^{-4\eta_*}) C_{\rm max}}{4}  \, \frac{\hbar a^4}{m} \,.
\label{K-deep:approx}
\end{eqnarray}
One of the simplifying assumption that we used to obtain the expression 
for $K^{(0)}(E)$ in Eq.~(\ref{K3-app}) was that $|s_{11}(x)| \ll 1$
for $x < 2.5$.  We can also use this assumption
to simplify the expression for $K_{\rm deep}(E)$
in Eq.~(\ref{K-deep:rl}).  The only dependence of $K_{\rm deep}(E)$
on the collision energy $E$ then comes from the factor $|s_{13}(x)|^2$ 
in the numerator.  The $^4$He results do not give significant constraints 
on this factor, because in the expression for $K^{(0)}(E)$ 
in Eq.~(\ref{K0:rl}), $s_{13}(x)$ appears 
in the combination  $s_{12}(x) s_{13}(x)$.
We will therefore make the simplifying assumption that
$|s_{13}(x)|^2$ is a sufficiently slowly varying function of $x$ 
for $x < 2.5$ that we can approximate $K_{\rm deep}(E)$ 
by its value at $E=0$, which is given in Eq.~(\ref{K-deep:approx}).

\begin{figure}[htb]
\centerline{\includegraphics*[width=12cm,angle=0,clip=true]{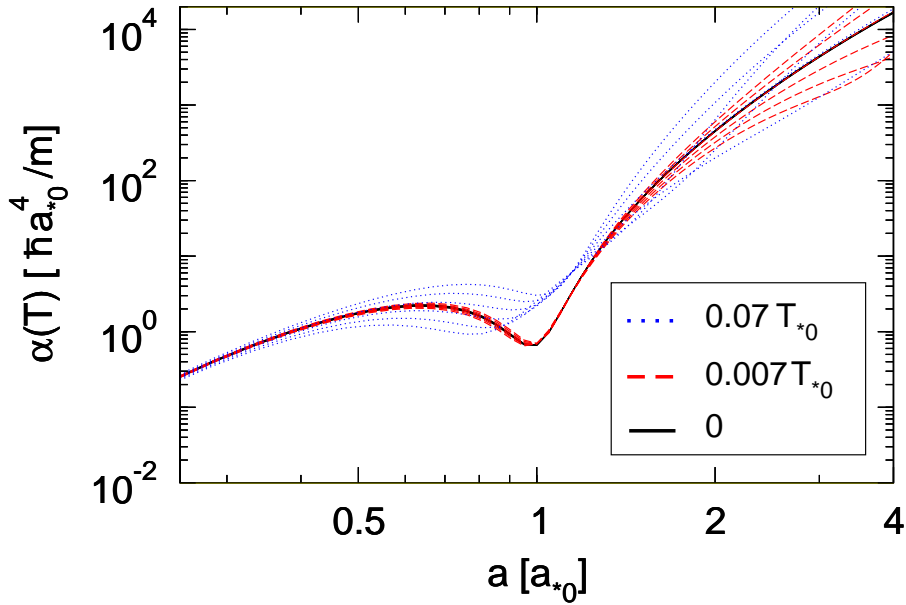}}
\vspace*{0.0cm}
\caption{The 3-body recombination event rate constant $\alpha(T)$ 
for atoms with deep dimers and a large scattering length.
The rate constant $\alpha(T)$ 
(in units of $\hbar a_{*0}^4/m$) is shown as
a function of $a$ for $\eta_*=0.01$ 
and several  values of the temperature $T$: 
$T = 0$ (solid line), 
$0.007 \, T_{*0}$ (dashed line), 
$0.07 \, T_{*0}$ (dotted line), 
where $k_B T_{*0} = \hbar^2/(m a_{*0}^2)$. 
The 6 curves for each temperature $T$ correspond to 
integer values of $t$ from $-2$ to $+3$.}
\label{fig:alphaT-T}
\end{figure}

\begin{figure}[htb]
\centerline{\includegraphics*[width=12cm,angle=0,clip=true]{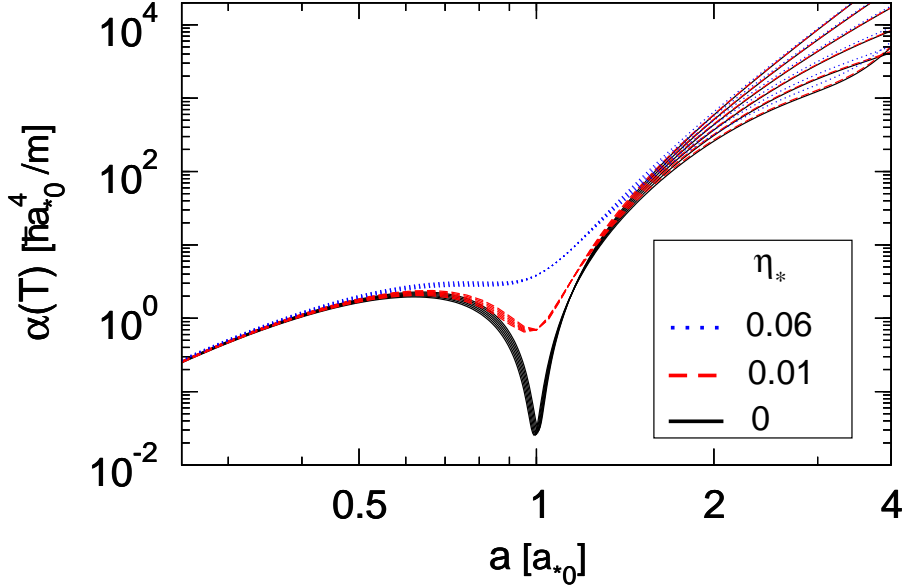}}
\vspace*{0.0cm}
\caption{The 3-body recombination event rate constant $\alpha(T)$ 
for atoms with deep dimers and a large scattering length.
The rate constant $\alpha(T)$
(in units of $\hbar a_{*0}^4/m$) is shown
as a function of $a$ for temperature $0.007 \, T_{*0}$, 
where $k_B T_{*0} = \hbar^2/(m a_{*0}^2)$,
and several  values of $\eta_{*}$: 
0 (solid line), 0.01 (dashed line), 
and 0.06 (dotted line).
The 6 curves for each value of $\eta_*$ correspond to 
integer values of $t$ from $-2$ to $+3$.}
\label{fig:alphaT-eta}
\end{figure}

The dependence of the 3-body recombination event rate constant 
$\alpha(T)$ on the scattering length $a$ is illustrated
in Figs.~\ref{fig:alphaT-T} and \ref{fig:alphaT-eta}.
The rate $K_3(E)$ in the numerator of Eq.~(\ref{alpha-T}) 
is the sum of $K_{\rm shallow}(E)$, for which the
$J=0$ term is given in Eq.~(\ref{K3-app:deep}) 
and the $J=1$, 2, and 3 terms are given in Eq.~(\ref{alpha-J:uni}),
and $K_{\rm deep}(E)$, which is approximated by Eq.~(\ref{K-deep:approx}).
For each value of $T$ and $\eta_*$, the range of predictions 
is shown by plotting 6 curves corresponding to integer values 
of the variable $t$ defined in Eq.~(\ref{h12}) ranging from $-2$ to $+3$.
The widths of the error bands reduce to zero at the $^4$He value 
$a=1.150 \, a_{*0}$.
In Fig.~\ref{fig:alphaT-T}, $\eta_*$ is fixed at the value 0.01
and $\alpha(T)$ is plotted as a function of $a$ 
for various values of $T$.
A convenient unit for the temperature is $T_{*0} = \hbar^2/(k_B m a_{*0}^2)$.
As $T$ increases from 0 to $0.007 \, T_{*0}$,
the prediction for $a$ near $a_{*0}$ is essentially unchanged.  
As $T$ increases further, the local minimum decreases in depth
and it tends to disappear around $T = 0.1 \, T_{*0}$.
In Fig.~\ref{fig:alphaT-eta}, $T$ is fixed at the value 
$0.007 \, T_{*0}$
and $\alpha(T)$ is plotted as a function of $a$ 
for various values of $\eta_*$.  As $\eta_*$ increases, 
the depth of the local minimum decreases until it
disappears around $\eta_* = 0.06$.

Our approximation for the 3-body recombination rate $K_3(E)$
depends on the fitted functions 
$h_1(x)$ and $h_2(x)$ in Eq.~(\ref{K3-app:deep}) and $f_J(x)$ in 
Eq.~(\ref{alpha-J:uni}).
If the only energies that contribute significantly 
to the Boltzmann average in Eq.~(\ref{alpha-T}) are ones that satisfy
$E \ll E_D$, we can use a much simpler approximation.  
We can approximate $K_3(E)$ by the sum of the $J=0$ term of 
$K_{\rm shallow}(E)$ at $E=0$, which is given in Eq.~(\ref{K-shallow:eta}), 
the leading term in the low-energy expansion of the $J=2$ term, 
which is given by Eqs.~(\ref{alpha-J:uni}) and (\ref{f2-x}), 
and $K_{\rm deep}(0)$, 
which is given in Eq.~(\ref{K-deep:approx}).
Inserting these expression into Eq.~(\ref{alpha-T}),
we get a simple expression for the rate constant for 3-body
recombination:
\begin{eqnarray}
\alpha_{\rm total}(T) \approx 
67.1 \left( \sin^2 [s_0 \ln (a/a_{*0}) ]
 + \mbox{$1\over2$} \cosh(2\eta_*) (1 - e^{-2\eta_*})   
 + 7.8 \, \left(\frac{k_B T}{E_D} \right)^2 \right) \, \frac{\hbar a^4}{m} \,.
\label{alpha-tot:approx}
\end{eqnarray}

\section{Application to $^{133}$Cs Atoms}

The Innsbruck group has carried out beautiful measurements of the 
3-body recombination rate for ultracold $^{133}$Cs atoms%
\footnote{A convenient conversion constant for $^{133}$Cs atoms
is $\hbar^2/m = 1.30339 \, {\rm K} \, a_0^2$.}
in the $|f=3, m_f=+3 \rangle$ hyperfine state \cite{Grimm06}.
By varying the magnetic field from 0 to 150 G, 
they were able to change the
scattering length from $-2500 \ a_0$ through 0 to $+1600 \ a_0$.
In this range of magnetic field, the $|f=3, m_f=+3 \rangle$ state
is the lowest hyperfine state, so 2-body losses are energetically 
forbidden.  Thus, the dominant loss mechanism is 3-body recombination.
The van der Waals length scale for Cs atoms is
$(mC_6/\hbar^2)^{1/4} \approx 200 \ a_0$.
The range of scattering lengths studied by the Innsbruck group 
includes a region of large negative $a$ and a region of large 
positive $a$ separated by a region of small $|a|$.
In the two regions of large scattering length, few-body physics 
should be universal.
An interesting open question is whether there is any relation
between the Efimov parameters $\kappa_*$ and $\eta_*$ that 
characterize the two universal regions.

In the region of negative $a$, the Innsbruck group measured the 
loss rate constant $L_3$ as a function of $a$ at three different 
temperatures: $T=10$ nK, 200 nK, and 250 nK.
They observed a dramatic enhancement of the loss rate for $a$ 
near $-850 \, a_0$.  At $T = 10$ nK, 
the loss rate as a function of $a$ can be fit 
rather well by the universal formula for $T=0$ in Ref.~\cite{Braaten:2003yc} 
with parameters $a_*' = -850(20) \, a_0$ and $\eta_* = 0.06(1)$.
Thus, the large enhancement in the loss rate can be explained by the 
resonant enhancement from an Efimov trimer near the 3-atom threshold.
More recently, the Innsbruck group has measured the position of the
maximum loss rate as a function of the temperature \cite{grimm1107}.
Its behavior as a function of temperature can be explained at least 
qualitatively by the dependence of the binding energy and width 
of the Efimov resonance on the scattering length \cite{Jonsell06,YFT06}.

\begin{figure}[htb]
\centerline{\includegraphics*[width=12cm,angle=0,clip=true]{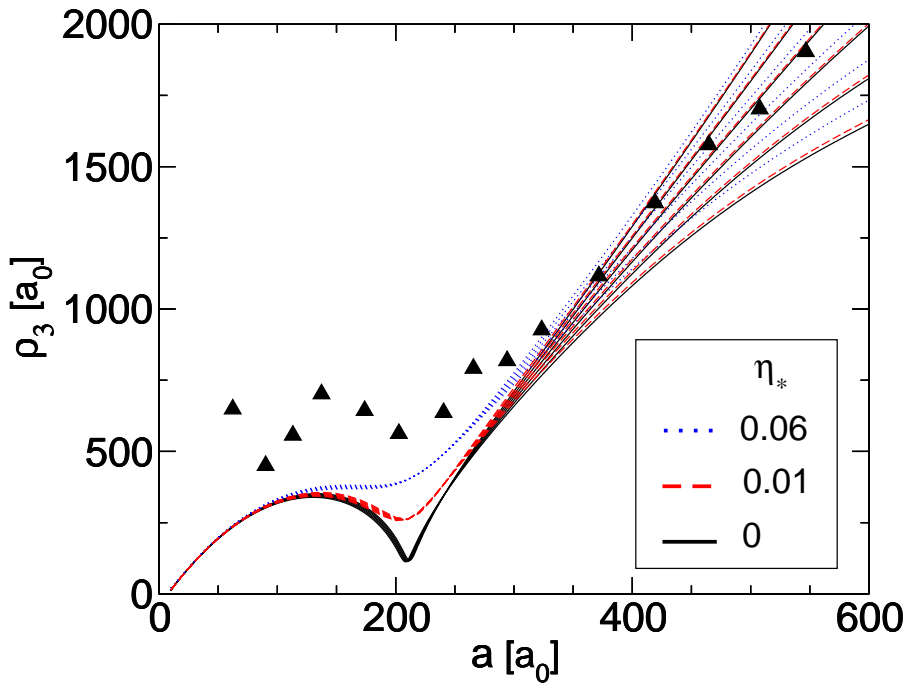}}
\vspace*{0.0cm}
\caption{The 3-body recombination length $\rho_3$ for $^{133}$Cs atoms 
as a function of $a$ for $T=200$ nK.
The data points are from Ref.~\cite{Grimm06}.
The curves are the universal prediction for several values 
of $\eta_*$: 
0 (solid lines), 0.01 (dashed lines), 
and 0.06 (dotted lines). 
The 6 curves for each value of $\eta_*$ correspond to 
integer values of $t$ from $-2$ to $+3$.}
\label{fig:alpha-Cs}
\end{figure}

In the region of positive $a$, the Innsbruck group measured the 
loss rate constant $L_3$ at $T=200$ nK for values of $a$ 
ranging up to $1228 \, a_0$.
Their results in the region $0 < a < 600 \, a_0$
are shown as solid triangles in Fig.~\ref{fig:alpha-Cs}.
The vertical axis is the recombination length $\rho_3$ defined by
\begin{equation}
\rho_3 = \left( \frac{2m}{\sqrt{3} \hbar} L_3 \right)^{1/4}.
\label{rho3}
\end{equation}
They observed a local minimum in the loss rate for $a$ 
near $200 \, a_0$.  This value is near the 
van der Waals length scale $(mC_6/\hbar^2)^{1/4} \approx 200 \ a_0$,
so there may be large corrections to the universal predictions.
The universal prediction for the loss rate at $T=0$ 
is $\alpha = K_3(0)/6$, where
$K_3(0)$ is the sum of Eqs.~(\ref{K-shallow:eta}) and (\ref{K-deep:approx}).   
By fitting the data for $a > 500 \, a_0$ to this
expression, they obtained $a_+ = 1060(70) \, a_0$ for the 
scattering length $e^{\pi/(2 s_0)} a_{*0}$
at which the coefficient of $a^4$ achieves its maximum value.  
Universality would then imply that the minimum 
should be at $a_{*0} = 223(15) \, a_0$.
The fit was insensitive to the value of $\eta_*$ and yielded 
only the upper bound $\eta_* < 0.2$.
The Innsbruck group also determined the location of the minimum directly 
by measuring the fraction of atoms that were lost after a fixed time.
The result was $a_{\rm min} = 210 (10) \, a_0$, which is consistent 
with the value obtained by fitting the data for $a > 500 \, a_0$.

We now consider whether our results obtained by scaling results from 
$^4$He atoms can be applied to this system of $^{133}$Cs atoms.
In the experiments with positive scattering length, the typical 
peak number density was $5 \times 10^{13} \, {\rm cm}^{-3}$.
The corresponding critical temperature for Bose-Einstein condensation
is $T_c \approx 160$ nK.  Thus, the temperature $T = 200$ nK
was not far above $T_c$.  We ignore this complication 
and calculate thermal averages using the Maxwell-Boltzmann distribution 
as in Eq.~(\ref{alpha-T})  instead of the Bose-Einstein distribution.
When $a = a_{\rm min}$, the universal prediction
for the binding energy $E_D$ of the shallow dimer in Eq.~(\ref{E-dimer})
gives $3 \times 10^4$~nK, 
which is about two orders of magnitude higher 
than the temperature.  The binding energy $E_D$ decreases to about
$3 \times 10^3$~nK at $a = 600 \, a_0$.  Thus the temperature $T = 200$~nK 
is safely in the region $T < 0.7 \, E_D$ in which the 
thermal average in Eq.~(\ref{alpha-T}) can be calculated accurately
from scaling the results for $^4$He atoms.

In Fig.~(\ref{fig:alpha-Cs}), we compare the universal predictions 
for the recombination length $\rho_3$ defined in Eq.~(\ref{rho3})
with the Innsbruck data.  
We assume $n_{\rm lost} = 3$, so that $L_3 = 3 \alpha$,
and we set $a_{*0} = 210 \, a_0$.  We plot $\rho_3$ for $T = 200$ nK 
as a function of $a$ for several values of $\eta_*$.
For each value of $\eta_*$, a band of predictions 
is obtained by plotting 6 curves corresponding to integer values 
of the variable $t$ defined in Eq.~(\ref{h12}) ranging from $-2$ to $+3$.
The band has zero width at $a = 242 \, a_0$, and the band remains 
narrow for $a$ near $a_{\rm min} = 210 \, a_0$.
The predictions fall well below the data in the region $a < 300 \, a_0$,
especially if $\eta_*$ is small enough that there is still a local minimum.
This suggests that $a_{\rm min}$ is too small for the universal 
predictions to be quantitatively accurate.
For $a > 500 \, a_0$, the scattering length is large enough that 
we expect the universal predictions to be valid. 
The results of Ref.~\cite{Grimm06} for $a > 400 \, a_0$
lie within the error bands of our universal predictions.
Unfortunately, the widths of the error bands are too large for 
$a > 500 \, a_0$ to allow us to determine the Efimov parameters
$a_{*0}$ and $\eta_*$.

\section{Summary}

In this work, we have used previously published results on the
3-body recombination rate of $^4$He atoms as a function of 
the collision energy to constrain the universal functions 
that govern 3-body recombination processes for atoms with large
scattering length. Since the scattering length for $^4$He atoms 
differs from the Efimov parameter $a_{*0}$ by only about 15\%,
the constraints are very strong near a minimum in the 3-body 
recombination rate as a function of scattering length. 
We were unable to fit the $^{133}$Cs data from the Innsbruck group 
near the local minimum of the recombination rate by adjusting the 
Efimov parameters $a_{*0}$ and $\eta_*$.  This is not a surprise, 
because the local minimum occurs at a value of $a$ that
is not large compared to the van der Waals length scale. 
The universal predictions should be accurate at larger values of the 
scattering length for which the recombination rate was measured by 
the Innsbruck group. Unfortunately the constraints from $^4$He atoms 
have large error bands at those larger values of $a$.
Our universal predictions obtained by scaling results for $^4$He atoms
can be applied to other systems of identical bosons for which 
the local minimum of the recombination rate occurs at a large 
value of the scattering length. 
For such a system, our results can be used to give a reliable determination 
of the Efimov parameters. 

In recent years, an effective field theory has been developed for
nonrelativistic few-body systems with large scattering 
length \cite{BHK99,BHK99b}.
It allows the model-independent calculation of few-body observables 
using a systematic expansion in $\ell/a$.  The leading term in this
expansion is the ``universal'' term.  This effective field theory  
can be applied to a large variety of physical
systems ranging from atomic physics to nuclear physics.
It  has been used to calculate the atom-dimer scattering 
phase shift as a function of the collision energy up to the 
dimer-breakup threshold \cite{BH02}.  The results were used to give universal 
predictions for the resonant dimer relaxation rate at nonzero temperature
\cite{Braaten:2006nn}.  The predictions were compared with 
measurements by the Innsbruck group, which revealed a resonant 
enhancement of the inelastic loss rate from a system of ultracold
$^{133}$Cs atoms and dimers \cite{FB18Santos}.  
This effective field theory can 
also be used to calculate the universal functions that determine the 
3-body recombination rate as a function of the collision energy.
Such a calculation would allow the Efimov parameters $\kappa_*$ 
and $\eta_*$ for $^{133}$Cs atoms to be determined by fitting 
the Innsbruck data for $a > 500 \, a_0$.  It could also be applied 
to any other system of identical bosons with large positive scattering length.
Once the Efimov parameters have been determined accurately,
it will be possible to make quantitative tests of the correlations 
between different aspects of Efimov physics that are predicted by 
universality. 

\begin{acknowledgments}
We thank H.-W.~Hammer for valuable discussions.
This research was supported in part by the Department of Energy 
under grants DE-FG02-05ER15715 (EB) and DE-FG02-93ER40756 (LP),
by an Ohio University postdoctoral fellowship
and by the Korea Research Foundation under grant 
KRF-2006-612-C00003 (DK).
\end{acknowledgments}

\end{document}